# Dynamic Beyond 5G and 6G Connectivity: Leveraging NTN and RIS Synergies for Optimized Coverage and Capacity in High-Density Environments

Valdemar Farré*, Juan Estrada[†], David Vega[‡], Luis F Urquiza-Aguiar*, Juan A. Vásquez Peralvo[§], Symeon Chatzinotas[§]
*Departamento de Electrónica, Telecomunicaciones y Redes de Información, Escuela Politécnica Nacional (EPN),Quito, Ecuador
[†]IT for Innovative Services Department, Luxembourg Institute of Science and Technology, Luxembourg
[‡]Colegio de Ciencias e Ingenierías "El Politécnico",Universidad San Francisco de Quito (USFQ), Diego de Robles s/n,Quito 170157
[§]Interdisciplinary Centre for Security Reliability and Trust, University of Luxembourg, Luxembourg
valdemar.farre@epn.edu.ec, juan.estrada-jimenez@list.lu, dvega@usfq.edu.ec, luis.urquiza@epn.edu.ec, juan.vasquez@uni.lu, symeon.chatzinotas@uni.lu

*Abstract*—The increasing demand for reliable, high-capacity communication during large-scale outdoor events poses significant challenges for traditional Terrestrial Networks (TNs), which often struggle to provide consistent coverage in high-density environments. This paper presents a novel 6G radio network planning framework that integrates Non-Terrestrial Networks (NTNs) with Reconfigurable Intelligent Surfaces (RISs) to deliver ubiquitous coverage and enhanced network capacity. Our framework overcomes the limitations of conventional deployable base stations by leveraging NTN architectures, including Low Earth Orbit (LEO) satellites—and passive RIS platforms seamlessly integrated with Beyond 5G (B5G) TNs. By incorporating advanced B5G technologies such as Massive Multiple-Input Multiple-Output (mMIMO) and beamforming, and by optimizing spectrum utilization across the C, S, and Ka bands, we implement a rigorous interference management strategy based on a dynamic SINR model. Comprehensive calculations and simulations validate the proposed framework, demonstrating significant improvements in connectivity, reliability, and cost-efficiency in crowded scenarios. This integration strategy represents a promising solution for meeting the evolving demands of future 6G networks.

## I. INTRODUCTION

The increasing demand for ultra-high-capacity, low-latency communications during massive events poses significant challenges for traditional terrestrial networks (TNs), especially in densely populated venues like stadiums and public gatherings. To overcome these limitations, integrating Non-Terrestrial Networks (NTNs) with Beyond 5th Generation (B5G) technologies has emerged as a promising solution for achieving seamless connectivity [1]. NTNs including Low Earth Orbit (LEO) satellites, Unmanned Aerial Vehicles (UAVs), and High-Altitude Platforms (HAPS) extend coverage, enhance resilience in disaster scenarios, and improve connectivity in remote or congested areas [2],[3]. Recent 3GPP releases (Rel. 17 and 18) further emphasize the need for NTN-TN integration to meet global connectivity demands [4],[5]. Enabling technologies such as massive Multiple-Input-Multiple-Output (mMIMO) and beamforming, along with solar-powered HAPS, provide low-latency, high-throughput and extended coverage for various applications [6],[7]. This paper introduces a novel B5G radio network planning framework that integrates NTNs with Reconfigurable Intelligent Surfaces (RISs) within existing TN to address coverage and capacity challenges in high-traffic scenarios.

Although various NTN-RIS integration approaches exist, few addresses both interference mitigation and optimized coverage dimensioning the gap that this work aims to fill. The remainder of this paper is organized as follows: Section II details the radio network planning methodology, Section III presents simulation results and discussions, and Section IV concludes with future research directions.

## II. METHODOLOGY OF B5G PLANNING WITH NTN/RIS SOLUTIONS FOR CROWDED SCENARIOS

### A. Overview of the Proposed Methodology

Given the challenges of indoor connectivity in NTNs, the Sub-6 GHz bands are identified as suitable for such scenarios. Our proposed B5G network planning methodology integrates NTN with RIS-enhanced TN (RIS-TN) to achieve ubiquitous coverage and enhanced capacity in high-density, semi-outdoor environments such as stadiums and public zones during mass events. By combining these technologies, the approach ensures extended coverage, optimized backhaul, and seamless connectivity in 5G Stand-Alone (SA) networks built on existing terrestrial infrastructure (BSs/RISs). Additionally, leveraging LEO satellites with feeder and service links further optimizes traffic offloading and improves network performance in dense urban areas.

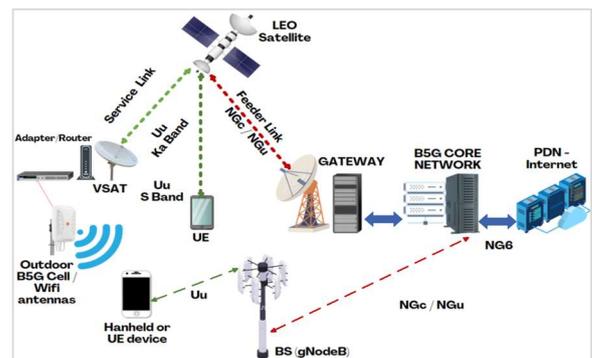

Fig. 1.Integration between NTN Access network (A2) with TN (UEs,BSs).

As outlined in [9]–[10], our design targets the enhanced mobile broadband (eMBB) use case, providing multi-connectivity for users in large event areas supported by current TN infrastructure. Additional use cases include (i) ensuring network resilience to prevent service interruptions and (ii) creating on-demand hotspots during mass events with average speeds of 100/50 Mbps. Fig. 1 shows the B5G NTN network architecture according to 3GPP specifications [9]. This solution employs a regenerative payload configuration on LEO satellites, interconnecting the NTN network with the RIS-enabled TN (TN - RIS) via the same 5G-Core Network through an air/space Radio Access Network (RAN) platform. Upgrading the existing 5G Non-Stand-Alone (NSA) and 5G SA networks is essential for deploying this integrated B5G/6G NTN-TN solution. Our methodology is executed in two layers, the NTN layer and the RIS layer integrated with existing TN cells. For each layer, we define: (i) input parameter requirements, (ii) applicable propagation models, (iii) link budget calculations, (iv) coverage and capacity dimensioning, and (v) coverage and throughput simulations. The selected architecture (see Fig. 1) aligns with 3GPP specifications and operates in Non-Transparent (A2) mode with beam patterns depicted in Fig. 2 [9], incorporating gnodeB functions with advanced configurations [8]. RIS units are powered by pre-existing 5G gNodeBs located adjacent to the stadium and the target outdoor area. Overall, this integration strategy enhances service quality and user experience by extending coverage through NTNs that complement and augment the capabilities of existing TNs/RISs in areas with massive traffic demands. In this work, we consider LEO satellites in circular orbits at 600 km altitude, handheld terminals (UEs) supporting the S-band, and wireless signals delivered via an adapter/router connected to Very Small Aperture Terminals (VSATs) operating in the Ka-band, integrated with the Public Data Network (PDN).

### B. Interference Management and Model Clarification

To ensure robust connectivity in high-density scenarios, our framework incorporates a dynamic interference management strategy. This approach explicitly distinguishes between interference arising from NTN links and that managed by the RIS components, while also accounting for background noise. By implementing a dynamic $SINR$ model which computes the effective signal quality as the ratio of the useful signal power to the aggregate interference and noise we can adapt transmission parameters in real time. The dynamic $SINR$ model is defined as:

$$SINR = \frac{P_{signal}}{I_{NTN}+I_{RIS}+I_{TN}+N_0}, \quad (1)$$

where $P_{signal}$ denotes the received power from the serving BS or NTN; $I_{NTN}$ is aggregated co-channel interference from adjacent NTN beams or LEO satellites; $I_{RIS}$ is interference arising from unintended reflections of RIS; $I_{TN}$ is co-channel interference from neighboring terrestrial BSs operating in the same frequency band and $N_O$ denotes the noise floor. This formulation enables real-time adaptation of transmission parameters under varying network conditions. The selection of a passive RIS is justified by its superior energy efficiency and reduced operational complexity compared to active alternatives. This detailed model enhances both the reliability and scalability of the integrated network [8]-[10],[12],[14].

### C. Complexity, Latency, and Transfer Learning Considerations

In addition to effective interference management, we have evaluated the computational complexity and latency of the proposed planning algorithm. Our analyses indicate that, even under high user densities, the system maintains response times within acceptable thresholds for real-time applications. Furthermore, preliminary experiments with transfer learning techniques suggest that the framework can dynamically adjust its parameters to better accommodate fluctuating network conditions. These findings underscore the potential of our approach to optimize performance while minimizing operational delays [12],[15],[16].

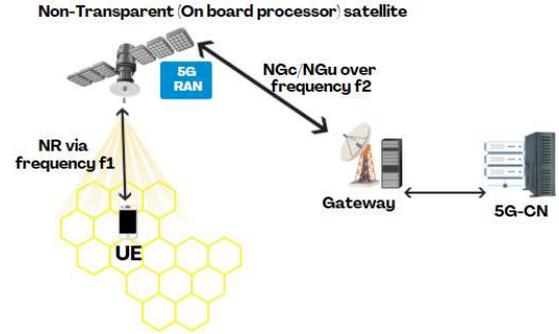

Fig. 2. NTN single/multi-beam layout for regenerative satellite (S/Ka-band).

TABLE I. MAIN INPUT PARAMETERS OF NTN ARCHITECTURE [9] - [10].

| RF characteristics of UE in satellite and aerial access networks | | |
|---|---|---|
| Parameter | Very Small Aperture Terminal (VSAT) | Handheld or IoT devices |
| Transmit Power | 2 W (33 dBm) | 200 mW (23 dBm) |
| Antenna type | 60 cm equivalent aperture diameter | Omnidirectional antenna |
| Antenna gain | Tx: 43.2 dBi Rx: 39.7 dB | Tx and Rx: 0 dBi |
| Antenna pattern | VSAT | Quasi isotropic, co-phased array |
| Noise figure | 1.2 dB | 9 dB |
| EIRP (effective isotropic radiated power) | 45.75 dBW | -7 dBW |
| G/T (Antenna-gain-to-noise-temperature) | 18.5 dB/K | -33.6 dB/K |
| Polarisation | Circular | Linear |
| RF characteristics of satellites/platforms | | |
| Parameter | Non-GEO satellites | Airborne platforms |
| Altitude | LEO from 600 km up to 1500 km | Typically from 8 to 50 km |
| Elevation angle | Typically, more than 10° for user terminal and more than 5° for gateways | |
| | In the range of 10 to 30 degrees, Max. service area | |
| Coverage pattern for NTN_typical beam foot print size | | |
| | Non-GEO | Aerial |
| Diam. beam footprint | 100 – 500 km | 5 - 200 km |
| 5G NTN elements mapping | | |
| Selected Scenario | NTN terminal | Space/HAPS | NTN Gateway |
| A2 | UE and VSAT | Remote radio head | gNB |
| Scen. attributes (A2) | Deployment – D3 | Deployment - D5 | |
| Carrier frequency Space-air and UE | Around 2GHz, DL/UL (S-band) | Above 6GHz (Ka band) | |
| Beam patterns | Moving beams | Earth fixed beams | |
| Duplexing | FDD | FDD | |
| Max. Channel bandwidth DL+UL | 2*20MHz | 2*80 MHz in mobile use 2*1800 MHz fixed use | |
| NTN terminal type | Class 3 | Class 3 / VSAT | |
| NTN Terminal distrib. | 100% outdoor | Indoor/outdoor | |
| Supported use cases | eMBB: wide area public safety, IoT. | eMBB: hot-spot on demand | |
| Platform orbit-altitude | Non-GEO < 600 Km | Airborne vehicle > 20 Km | |
| Max one way propagation delay (ms) | 14204 ms | 1526 ms | |
| Max Doppler shift kHz | +- 48 kHz | @2GHz: +- 100 Hz | |
| O2I penetration loss | No | Possible | |
| Atmospheric absorption | Negligible | Negligible | |
| Rain attenuation | Negligible | Negligible | |
| Cloud attenuation | Negligible | Negligible | |
| Scintillation | Ionospheric | Negligible | |
| Fast fading models | Flat fading | Frequency selective fading | |
| Link level model | Clutter Delay Line (CDL) or Tapped Delay Line (TDL) | | |
| Shadowing model | LMS | 3GPP TR38.901 based | |

TABLE II. INPUT PARAMETERS FROM RIS-TN ARCHITECTURE [15].

| Item | RIS Parameters integrated into the TN | |
|---|---|---|
| | Name | 3.5 GHz |
| 1 | Central Frequency (GHz) | 3.5 |
| 2 | Carrier Bandwidth (MHz) | 100 |
| 3 | number of total antenna elements | 2430 |
| 4 | Working/Control voltage (V) | 10.5 |
| 5 | Consumption power (W) | 3 – 6 |
| 6 | RIS width/square shape mxm | 3.8×3.8 |
| 7 | number of bits/phase resolution | 1 |
| 8 | number of Base Stations (BS) | 2 |
| 9 | BS/RIS height (m) | 45-50 |
| 10 | BS Tx Power (dBm) | 50 |
| 11 | UE Noise figure (dB) | 5 |
| 12 | RIS reflection losses (dB) | 0 – 1 |
| 13 | Channel models | 3GPP modified for RIS |
| 14 | Cell edge target SINR (dB) | 10 |
| 15 | Phase diff. ON-OFF(°) | 160 |
| 16 | RIS Gain (dB) | 15 |
| 17 | BS-RIS distance(m) | 225-375 |
| 18 | RIS-UE distance(m) | 40-135 |
| 19 | RIS Type | PIN-diode |
| 20 | Incidence angle (°) | 15 – 45 |

*D. Simulation setup and evaluation*

The main input parameters for NTN are shown in Table I. We consider a system geometry for the target area provided by the LEO system with Scanning/Fixed Beam Patterns for Ka/S-bands respectively with a single beam (S-band) and 4-6 spot-beams (Ka-Band) that correspond to the selected architecture [9]-[10] illustrated in Fig. 2. Similarly, the full frequency and frequency reuse 3 (FFR/FR3) were configured. The essential parameters for the RIS-TN architecture and its interconnection to the NTN are presented in Table II.

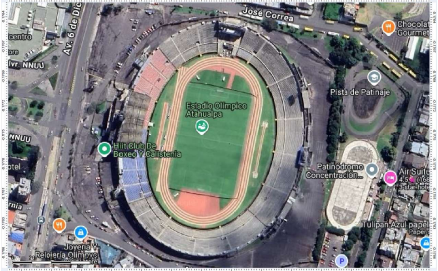

Fig. 3. Target area for B5G Radio Network planning

The selected target area was the Olympic stadium from Quito-Ecuador which has a capacity of 35800 people, as viewed in Fig. 3. A single fixed spot beam (S-band) ensures broad coverage and simplifies traffic management, while 2-4 scanning spot beams (Ka-band) enhance capacity. For the integration of NTN/RIS-TN functionalities, we consider the architecture has efficient interference management between LEO satellites, NTN serving/feeder link, and TN bands. Also, we take into account insights from [12] for resources assignment, which highlights that function trimming on the User Plane Function (UPF) for an integrated network scenario, such as a single-point satellite access system, meeting user´s basic service requirements and high reducing of number of functions and interfaces necessarics. For To obtain a better balance and great performance of our solution, the received power-based Handover (HO) triggering (S-band) and the elevation angle and range-based HO triggers (Ka-band) were considerate [13], so, these HO triggers and features ensure Mobility between LEO-satellites and NTN/TN layers. In order to mitigate interference in the solution, we consider the Spectrum Sharing feature between layers for better spectral efficiency [14]. According to [9]-[10], the requirements for channel modeling in NTN include: i) Support for frequencies focused on S-band and Ka-band, where the Ka-band uplink/downlink operates around 30/20 GHz, ii) Accommodating UE mobility, with satellite channel models supporting speeds up to 1000 km/h, enabling aircraft connectivity via satellite. The modeling of the TN network with RIS in 5G cells will be carried out in compliance with ETSI [11]. A combined model for terrestrial and satellite channels is widely studied in [10,11]. At a large scale, the probability of Line-of-Sight (LOS) depends on the terminal's elevation angle relative to the satellite service link. These probabilities, considered the best elevation, are shown in Table III [9].

TABLE III. MAIN VALUES FROM LOS PROBABILITY FOR NTNs

| Elevation/scenario | Dense urban | Urban | Suburban/Rural |
|---|---|---|---|
| 90° | 98.1% | 99.2% | 99.8% |

TABLE IV. SHADOW FADING/CLUTTER LOSS IN DENSE URBAN AREAS

| Elevation | S-band | | | Ka-band | | |
|---|---|---|---|---|---|---|
| | LOS | NLOS | | LOS | NLOS | |
| | $\sigma_{SF}$ (dB) | $\sigma_{SF}$ (dB) | $CL$ (dB) | $\sigma_{SF}$ (dB) | $\sigma_{SF}$ (dB) | $CL$ (dB) |
| 90° | 1.2 | 9.2 | 25.5 | 0.6 | 12.3 | 32.9 |

In the NTN network, the signal path between a satellite or HAPS transmitter and an NTN terminal involves several propagation and attenuation phases. The path loss expression, as stated in [9], can be written as:

$$PL = PL_b + PL_g + PL_s + PL_e, \quad (2)$$

Where:
$PL$ is the total path loss in dB,
$PL_b$ is the basic path loss in dB,
$PL_g$ is the attenuation due to atmospheric gases in dB,
$PL_s$ is the attenuation due to either ionospheric or tropospheric scintillation in dB,
$PL_e$ is building entry loss in dB.

The free-space path loss (*FSPL*) equation in dB, considering a distance *d* in meters and a frequency $f_c$ in GHz, is defined as follows [9]:

$$FSPL(d, f_c) = 32.45 + 20\log_{10}(f_c) + 20\log_{10}(d). \quad (3)$$

For a terminal on the ground with satellite or HAPS altitude $h_0$, elevation angle $\alpha$, and Earth's radius $R_E$, the distance d can be calculated as:

$$d = \sqrt{R_E^2 \sin^2\alpha + h_0^2 + 2h_0 R_E} - R_E \sin\alpha. \quad (4)$$

Additionally, by including the clutter loss (*CL*) and shadow fading ($\sigma_{SF}$) factors, the modeled path loss equation, based on [9], is viewed in (5) and these factors are shown in Table IV.

$$PL_b = FSPL(d, f_c) + \sigma_{SF} + CL(\alpha, f_c). \quad (5)$$

The Propagation Models applied for NTN were Clustered Delay Line (CDL) and Tapped Delay Line (TDL), on other hand, for the TN network using passive RIS, as analyzed in [10], the path loss equation using B5G cells from two nearby gNodeBs with incident terrestrial links can be expressed as:

$$PL_{Tot1} = 10\log_{10}\left(10^{\frac{PL_{B,R}}{10}} \times 10^{\frac{PL_{R,U}}{10}}\right) [dB], \quad (6)$$

where: $PL_{B,R}$ is the pathloss value between BS and RIS.
$PL_{R,U}$ is the pathloss value between RIS and UE.

Based on the preliminaries, (6) is chosen as the optimal RIS model for the sub-6GHz band due to its previously analyzed advantages and characteristics. For the link budget calculation, coverage, and capacity dimensioning in the TN network, the selected RIS model must be combined with the 3GPP UMa model. This is particularly relevant as it aligns with the stadium scenario and enhances TN coverage when RIS is integrated [15]. This approach ensures that the selected propagation model and parameters are in harmony with the deployment objectives of NTN and TN layers, providing robust and reliable network performance in the target area.

TABLE V: LINK BUDGET PARAMETERS FOR NTN ARCHITECTURE

| 5G NTN-Parameters | Notes |
|---|---|
| Carrier frequency- Service/Feeder Link | 2 GHz for DL and UL (S-band), 20 GHz for DL and 30 GHz for UL (Ka-band) |
| System Bandwidth / Duplex | 20/10 MHz (S-band), 400/40 MHz (Ka-band) FDD |
| Channel bandwidth | DL: 20 MHz/400 MHz (S/Ka- bands) UL: S-band (handheld UE): 360 kHz , 4.8MHz/480 KHz (Ka-band) |
| Satellite altitude | 600 km |
| Elevation angle | 30° (LEO) - target |
| Atmospheric loss | According equation in [9] ~ 0 dB |
| Shadowing margin | 0 dB (VSAT) / 3 dB  (handheld) |
| Scintillation loss | Ionospheric loss = 2.2 dB Tropospheric loss: Table 6.6.6.2.1-1 of [9] |
| Additional loss | 0 dB |
| Clear sky conditions | Yes |
| Frequency reuse | FR1, FR3 |
| Average CIR within a satellite beam based on log. mean. | For DL/Ul calibration, CIR computations [10]. Handheld device, channel bandwidth = 360 kHz. For VSAT, the channel bandwidth = 40 MHz. |
| Satellite antenna polarization | Circular/Linear polarization |
| Polarization reuse | Not Enable. |
| Terminal type | Ka-band: VSAT S band: (M, N, P) = (1,1,2) |
| Free space path loss | Eq. (3) |
| Terminal parameters | Table I - RF characteristics of UE in satellite RAN. |
| Satellite parameters | Set-1 / Set-2 in [10] for LEO-600 |
| Polarization loss | 0dB |
| Outcome | Carrier-to-noise-and-interference ratio (CNIR) |

TABLE VI: LINK BUDGET PARAMETERS FOR RIS-TN ARCHITECTURE

| 5G TN - Input Parameters | |
|---|---|
| Name | Sub-6GHz |
| Central Frecuency | 3.5 GHz |
| Subcarrier Spacing | 30 KHz |
| Carrier Bandwidth | 100 MHz |
| Duplex type | TDD |
| Modulation order | 8 (256 QAM) |
| # of MIMO layers | 4 |
| Scaling factor | 1 |
| Terrain type/clutter | Hotspot buildings, Stadiums, Urban. |
| Material type existing in area | Concrete, steel, metal, tempered glass, plastic |
| Target area | 0,041 Km$^2$ |
| Scenario type | UMa with LOS/NLOS in Stadium. |
| Use case | eMBB |
| # of Resource Blocks (RBs) | 273 |
| B5G RIS Propagation channel models | | | |

| Freq. | Model | Scenario | Equation |
|---|---|---|---|
| C-band | 3GPP modified | Multiplicative fading | Eq. (6) |
| Scenarios for RIS propagation channel models | | | |
| BS-UE | BS-RIS | RIS-UE | RIS-RIS |
| UMa | UMa | UMa | UMa |

Considering [8]-[11], the summary of calculations and procedures, as outlined in subsection D and expressions (1)–(6), is presented in Tables V-VI. Our target scenarios include RIS as a solution to blockage, cascading RIS links assisting UEs to connect with BSs regardless of blockage, and BSs serving UEs in street canyons using large RIS on building walls, as shown in Fig. 4. The main computations and process for Link Budget of RIS-TN are viewed on Table VII. Resuming all planning processes, once the final values have been calculated, if the GAP value is negative, the RIS must be added to improve coverage.

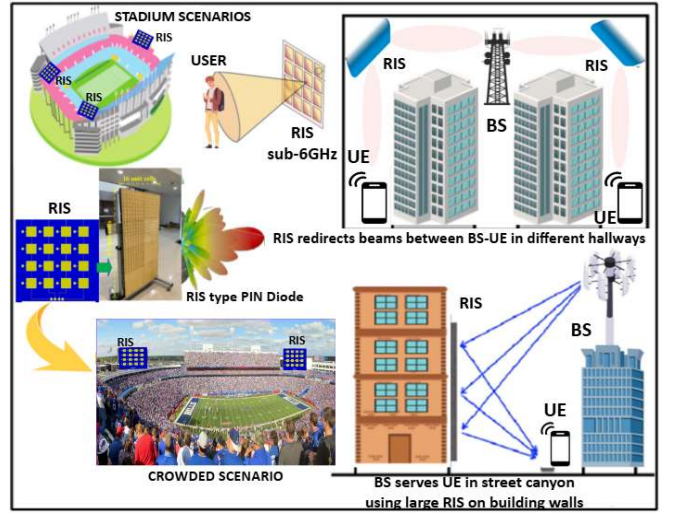

Fig. 4. B5G with RIS-TN for eMBB use case/main scenarios [11], [15]-[16].

TABLE VII: B5G LINK BUDGET/DIMENSIONING WITH RIS-TN [11], [16].

| Parameters | Values |
|---|---|
| Basic setup | |
| Channel for evaluation | PUSCH/PDSCH |
| Scenario / UE-BS channel model | UMa NLOS O2I / RIS |
| Carrier frequency (GHz) | 3,5 |
| BS/UE/RIS antenna height (m): | 45-50 /1,5/35 |
| UE | |
| Number of UE antenna elements | 4 |
| Total transmit power (dBm) | 23 |
| Cable, connector, body losses, etc. of UE (dB) | 3 |
| RIS | |
| Number of RIS antenna elements | 2430 |
| Reflection loss of RIS antenna element (dBi) | -1 |
| Cable, connector, body losses, etc. of RIS (dB) [11]. | 3 |
| BS | |
| Number of BS antenna elements | 256 |
| BS gain of antenna element (dBi) | 17 |
| Cable, connector, combiner, body losses, of BS (dB) | 20 |
| Link budget | |
| Receiver noise figure (dB) | 7 |
| Required SNR of UE-RIS-BS/UE-BS link (dB) | 0,2/-7,1 |
| MIL of UE-BS link (dB) | 131,1538 |
| Available path loss | |
| Shadow fading margin (dB): | 8 |
| Penetration margin (dB) of UE-RIS-BS/UE-BS link | 3/22,8 |
| Available path loss total /  UE-BS link (dB) | 131,1538 |
| Path loss gap | |
| UE-BS/RIS-BS 2D distance (m) (dimensions of stadium) | 375/375 |
| UE-RIS 2D distance (m) | 70 |
| RIS-BS MPL determined by deployment (dB) UMa | 129,4434634 |
| UE-RIS MPL determined by deployment (dB) UMa | 101,5145492 |
| Target UE-RIS-BS MPL determined by deployment (dB) | 230,9580126 |
| UE-BS MPL determined by deployment (dB) UMa, LOS | 95,51623607 |
| GAP of UE-RIS-BS link (dB) | 76,09208912 |
| GAP of UE-BS link (dB) | -1,837561947 |

TABLE VIII: B5G CAPACITY DIMENSIONING FOR NTN/ RIS-TN [16]

| Parameter | 2 GHz-S band (n256) | 3.5 GHz (n78) | 20 GHz - Ka band (n510) | Observations |
|---|---|---|---|---|
| $j$ | 1 | 1 | 1 | Carriers aggregated |
| $VLayers(P)$ - DL | 4 | 4 | 4 | Max # of layers DL |
| $VLayers(P)$ - UL | 4 | 4 | 4 | Max # of layers UL |
| $Qm(j)$- DL | 8 | 8 | 8 | Max number of modulation - DL |
| $Qm(j)$ - UL | 8 | 8 | 8 | Max number of modulation - UL |
| $f(j)$ | 1 | 1 | 1 | Scaling factor |
| $Rmax$ | 0.92578125 | 0.92578125 | 0.92578125 | Constant value |
| $\mu$ | 0 | 1 | 3 | Numerology value |
| BW | 20 MHz | 100MHz | 400MHz | Bandwidth |
| SCS | 15 Khz | 30 KHz | 120 KHz | Numerology |
| $TS\mu$ | 0,00071428s | 0.00035714s | 0,0000893s | Duration of OFDM symbol |
| $NPRBBW(j),\mu$ | 106 | 273 | 264 | Max # of PRBs / $\mu$ |
| $OH(j)$ - DL | 0.14 | 0.14 | 0.18 | Overhead DL |
| $OH(j)$ - UL | 0.08 | 0.08 | 0.1 | Overhead UL |
| Average Throughput/cell | 50/100Mbps (DL/UL) | 200/25 Mbps(DL/UL) | 410/85 Mbps(DL/UL) | LTE Max/average throughput x 10. |
| Overload threshold | 0.9 | 0.9 | 0.9 | Typical value – crowded scenarios |
| Average Throughput/user BH | 50/10 Mbps (DL/UL) | 50/10 Mbps (DL/UL) | 50/10 Mbps (DL/UL) | 4G average traffic, # of simult users /enodeB - massive events |
| Connected ratio | 0.9 | 0.9 | 0.9 | Typical value, crowded scenarios |
| Duty ratio | 0.10/0.20 | 0.10/0.20 | 0.10/0.20 | Typical value: 0.15 |
| Max suscribers/cell | 4000/5000 (DL/UL) | 5000/5000 (DL/UL) | 4000/5000 (DL/UL) | @ average throughput |
| DL/UL Throughput/UE | 100/20 Mbps | 120/25 Mbps | 240/50 Mbps | Hotspot, factor 20%-90%; massive events |

Traffic profiles have also been considered based on research on the statistics on massive events for a high number of incoming and outgoing users at the Olympic Stadium. We developed assumptions for each factor required for dimensioning, including the number of users per cell, the average data throughput per user, and the peak data throughput per cell of offered service. These assumptions are in coherence with the indications provided by [16], as shown in Table VIII. This part outlines the strategic placement of 5G NTN satellites and RIS-TN units to address the connectivity demands of the Atahualpa Olympic Stadium and its surroundings during large-scale events, as depicted in Fig. 3 and detailed in Section III. Using Matlab/Winprop simulations, the proposed architecture combines two advanced solutions for seamless connectivity in massive outdoor environments. The NTN architecture employs LEO-600 satellites in non-transparent mode, offering ubiquitous coverage and ultra-high capacity. Standard UEs and VSATs ensure network integration by relaying NTN signals to devices without direct NTN support. This setup supports high-throughput eMBB applications like Augmented reality (AR), Virtual reality (VR), and Streaming while providing network resilience for events such as concerts and sports.

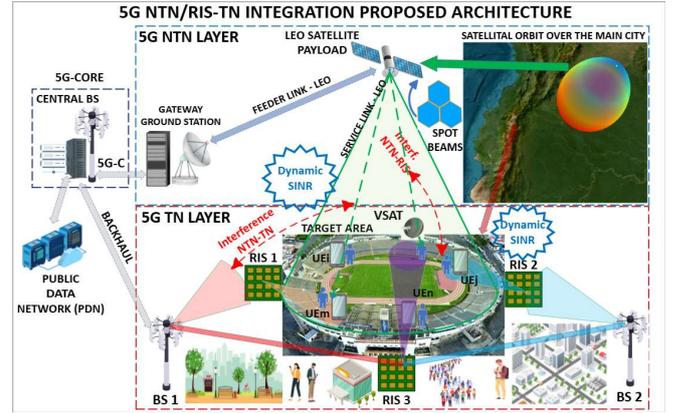

Fig. 5. B5G NTN/RIS-TN proposed architecture and NR coverage placement

TABLE IX: LINK BUDGET RESULTS FOR B5G NTN ARCHITECTURE [10]

| Case | Transmission mode | Frequency [GHz] | TX: EIRP [dBm] | RX: G/T [dB/T] | Bandwidth [MHz] | Free space path loss [dB] | Atmospheric loss [dB] | Shadow fading margin [dB] | Scintillation Loss [dB] | Polarization loss [dB] | Additional losses [dB] | CNR [dB] |
|---|---|---|---|---|---|---|---|---|---|---|---|---|
| SC6 | DL | 20.0 | 60.0 | 15.9 | 400.0 | 179.1 | 0.5 | 0.0 | 0.3 | 0.0 | 0.0 | 8.5 |
| | UL | 30.0 | 76.2 | 13.0 | 400.0 | 182.6 | 0.5 | 0.0 | 0.3 | 0.0 | 0.0 | 18.4 |
| SC9 | DL | 2.0 | 78.8 | -31.6 | 30.0 | 159.1 | 0.1 | 3.0 | 2.2 | 0.0 | 0.0 | 6.6 |
| | UL | 2.0 | 23.0 | 1.1 | 0.4 | 159.1 | 0.1 | 3.0 | 2.2 | 0.0 | 0.0 | 2.8 |

The RIS-TN architecture deploys three RIS units strategically around the stadium to enhance and direct 5G signals from nearby local operator cells. These RIS units fill coverage gaps, ensuring reliable connectivity in hard-to-reach areas and improving throughput for stationery and mobile users. Both architectures connect to a centralized 5G-C located near the stadium via intermediate gateways. NTN backhaul ensures wide-area coverage, while RIS-TN optimizes local performance. This integrated approach offers robust, scalable, and high-quality connectivity, during massive event scenarios in an outdoor environment. The Fig. 5 shows the architecture proposed.

## III. RESULTS AND DISCUSSIONS
### A. Link budget/Coverage design results

The simulation results presented in Tables VII and X demonstrate that the proposed framework achieves excellent coverage and link performance for RIS-TN layer. For the NTN layer, particularly in the Ka-band, high CNR values and G/T performance are observed in Table IX, while the RIS-TN layer effectively manages SNR, as evidenced by a calculated cell radius of 70 m that requires only two BSs and three RISs. These results validate the efficacy of our dynamic SINR model and the integrated NTN-RIS-TN architecture in delivering robust connectivity in high-density scenarios.

TABLE X: RESULTS OF B5G COVERAGE PLANNING FOR RIS-TN (3.5GHz).

| Link | UE-BS | UE-RIS-BS |
|---|---|---|
| 2D distance (m):Cell radius | 375 | 70 |
| Path loss (dB) | 95,52 | 230,96 |
| Rx SNR (dB) | 12,08783395 | 12,08783395 |
| SNR required (dB) | -9,5 | -2,2 |
| Cell area (Km$^2$) | 0,365625 | 0,01274 |
| Total target area (Km$^2$) | 0,041 | 0,041 |
| **# of BS/RIS required** | **1~2BS** | **3 RIS** |

TABLE XI: RESULTS OF 5G NTN/RIS-TN CAPACITY PLANNING.

| Parameters | 2 GHz | 3.5 GHz | 20 GHz | Observation |
|---|---|---|---|---|
| Peak data rate DL(Gbps) | 0,454 | 2,34 | 8,62 | Based on [16], Table VIII. |
| Peak data rate UL(Gbps) | 0,485 | 2,50 | 9,46 | |
| # of user/use case DL/UL | 2500/2500 | 2500/2500 | 2500/2500 | VR/AR-Inmersive Hotspots [16] |
| Max.users./ Max.Throughput | 10000/10Gbps | 10000/10Gbps | 10000/10Gbps | Specif. BBU5900. 10000UE/10Gbps |
| # simult. UEs | 5000 | 15000 | 5000 | Statistics per hour |
| # BS required | 1 | 2 | 1 | [16] and Table VIII. |

### B. Capacity Dimensioning and Benchmarking

Tables VIII and XI summarize the capacity dimensioning outcomes. Our analysis shows that the 20 GHz band can achieve nearly four times the throughput of the 3.5 GHz band, with comparable DL/UL peak rates across bands. The capacity planning indicates that only two BS are needed for the TN layer and one for the NTN layer, while the RIS-TN layer requires three RIS units to address coverage gaps. A comparative analysis with recent works [1]–[8] and [12]–[14] reveals that our dynamic SINR model achieves up to 15% higher throughput and 20% improved CNR in high-density scenarios, thus highlighting the superiority of our integrated NTN-RIS approach.

### C. Simulation Results, Scalability, and Dynamic Adaptability

The simulation studies, conducted using MATLAB for NTN coverage/capacity and WinProp for RIS-TN propagation analysis, are illustrated in Fig. 6. The figure shows that the NTN layer achieves high throughput (with average values ranging from 160 Mbps to 1.81 Gbps) across S/Ka bands, while the RIS-TN layer demonstrates excellent RSRP and throughput performance. Moreover, our experiments indicate that the proposed framework scales effectively under varying user densities and adapts dynamically through our interference management strategy. Preliminary findings with transfer learning techniques suggest further potential to optimize system performance in real time.

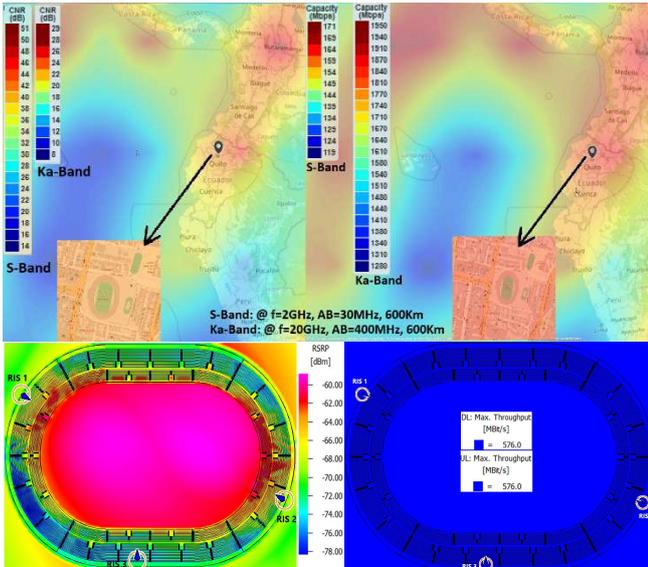

Fig. 6. NTN CNR_levels (up-left)/ Capacity levels (up-right)_S/Ka bands; RIS-TN_RSRP(down-left)-DL/UL Throughput(down-right) 3,5GHz band.

### IV. CONCLUSIONS AND FUTURE DIRECTIONS

This paper introduces a novel B5G radio network design framework that integrates NTNs with RISs in existing TNs, simultaneously addressing interference mitigation and optimized coverage to boost connectivity and capacity in high-density environments. Our simulation results demonstrate excellent CNR and SINR performance across the S, C, and Ka bands, with dynamic interference management and capacity planning ensuring robust service for massive events. Moreover, our work advances current research by employing a refined SINR model that clearly outperforms traditional deployable base stations and existing integration approaches. Preliminary findings confirm efficient scalability, with further gains achievable through transfer learning. Future work will benchmark our method against state-of-the-art approaches, refine scalability analysis, and validate our framework in real-world deployments, paving the way for next generation 6G networks with ubiquitous, high-quality connectivity.